\documentclass[showpacs,preprintnumbers,titlepage,preprint,showkeys,nofootinbib,bibnotes,letterpaper,10pt,balancelastpage,twocolumn,aps]{revtex4}
\usepackage{amsmath}
\usepackage{graphicx}
\usepackage{dcolumn}
\usepackage{bm}

\begin{document}

\title{Different thermal conductance of the inter- and intra-chain
interactions in a double-stranded molecular structure}
\author{Wei-Rong Zhong}
\email{wrzhong@jnu.edu.cn}
\affiliation{\textit{Department of Physics, College of Science and Engineering, Jinan
University, GuangZhou 510632, P. R. China}}
\date{\today }

\begin{abstract}
A double-stranded system, modeled by a Frenkel-Kontorova lattice, is studied
through nonequilibrium molecular dynamics simulations. We have investigated
the thermal conductance influenced by the intra-chain interaction as well as
by the inter-chain interaction. It is found that the intra-chain interaction
always enhance the thermal conductance. The inter-chain interaction,
however, has a positive effect on the thermal conductance in the case of
strong nonlinear potential, and has a negative effect on the thermal
conductance in the case of weak nonlinear potential. This phenomenon can be
explained by the transition of thermal transport mode and the phonon band
shift of the particles. It is suggested that the inter-and intra-chain
interactions present different thermal properties in double-stranded
lattices.
\end{abstract}

\maketitle

\textbf{I. INTRODUCTION}

Deriving macroscopic physics laws from simple microscopic models is one of
the tasks of non-equilibrium statistical mechanics \cite{Lepri1}. So far
single-chain lattices have attracted great interest and extensive studies in
the recent decades for the simple reason that they are easier to study
through simulations and through whatever analytical methods are available
\cite{Pereira} \cite{Lebowitz} \cite{Hu} \cite{Zhong}. For one-dimensional
single-chain systems, some of the most interesting results that have been
obtained are as below: 1) the heat current $J$ decreases with system size $N$
as $J\symbol{126}1/N^{\alpha }$, where $\alpha <1$ \cite{Dhar} \cite{Narayan}
\cite{Lepri2} \cite{Nadler}. 2) the thermal conductivity $\kappa $ increases
with the inter-particle coupling $\lambda $ as a scaling law $\kappa \symbol{%
126}\lambda ^{\gamma }$, here $\gamma \symbol{126}1.5$ \cite{Shao}.\ In
other words, the thermal conductivity is directly proportional to the
interaction of the particles in single-chain systems.

In double-stranded and multi-stranded lattices, which are more common
structures in nature such as DNA double helix \cite{Wolfram}, double- and
multi-chain polymers \cite{Brandrup}, multi-stranded nanofibers \cite%
{Maheshwari} and nanotubes \cite{Connell} \cite{Diao}, there are few
detailed studies and it is fair to say that it is totally unclear as to
whether they have the same properties as those the single-chain lattices
have, and if not, then what they are different from. Liu and Li \cite{Liu}
have studied the inter-chain coupling in simple networks consisting of
different one dimensional nonlinear chains. They reported that the coupling
between chains has different functions in heat conduction comparing with
that in electric current. However, the double-chains they have studied,
which are coupled together only with two particles, are far different from
the true double-stranded structure. For two dimensions nonlinear systems,
which are much similar to double-stranded and multi-stranded lattices, Lee
and Dhar \cite{Lee} have performed simulations to determine the system size $%
(L)$ dependence of the heat current $(J)$. Unfortunately, the inter-and
intra-chain interactions in their models have been still regarded as the
same interactions.

Up to now lots of the studies about the low dimensional thermal conductance
focus on the single-chain lattices and the couplings of the particles \cite%
{Lepri1}, which is called the intra-chain interactions here. However, few
researchers address on the influence of the inter-chain interactions on the
thermal conductance. Moreover, in biological and chemical research areas,
some previous investigations have shown that inter-chain interactions play
an important role in the properties of materials. For instance, the
inter-chain interactions can affect the denaturation of DNA \cite{Huang}
\cite{Nelson} \cite{Drukker} \cite{Yu}, the luminescent properties of
polymer \cite{Wu} and the electronic properties of the dimer \cite{Ottonelli}%
. Therefore, it is\ highly necessary to study the thermal properties and
relevant transport problems of the inter-chain interaction. Particularly,
the difference between the intra and inter-chain interaction is an important
issue to be discussed. 

\begin{figure}[htbp]
\begin{center}\includegraphics[width=8cm,height=6cm]{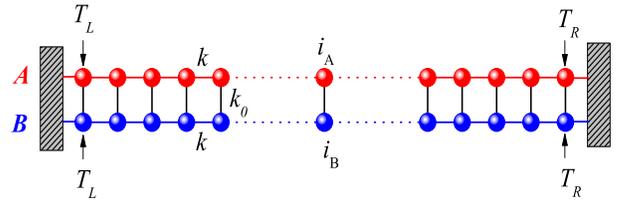}
  \end{center}
  \caption{Diagram
of the double-stranded lattices. $A$-chain (red) and $B$-chain (blue) are
coupled together via a harmonic spring $k_{0}$. $k$ is the strength of
intra-chain interaction. $T_{L}$ and $T_{R}$ are the temperature of the heat
baths connected to the first and the last particles of each chain.}
   \label{}
\end{figure}

In this paper, We consider heat conduction in a double-chains
Frenkel-Kontorova (FK) system as shown in Fig. 1. We expect\ that the
inter-chain interaction can lead to some interesting and rather surprising
results in thermal conductance. In this model the inter-chain interaction
means the interaction between $A$-chain and $B$-chain. The intra-chain
interaction, which is corresponding to the coupling of the particles in
single chain system, refers to the coupling of the particles in $A$-chain or
in $B$-chain. It have been reported that strong intra-chain interactions
always raise the heat conductance in the single-chain FK lattice \cite%
{Lepri1} \cite{Shao}. A similar calculation for the inter-chain interaction
is so far not available in the double-chains case, and we will address this
specific question.

\textbf{II. MODEL AND SIMULATION METHOD}

We consider heat conductance in a double-stranded FK crystal described by a
total Hamiltonian

\begin{equation}
H=H_{A}+H_{B}+H_{int}.
\end{equation}%
As shown in Fig.1, we couple the $i$th particle of $A$-chain with the same
order particle of $B$-chain via a harmonic spring, and the coupling
Hamiltonian $H_{int}=\sum_{i=1}^{N}[k_{0}\left( x_{A,i}-x_{B,i}\right) ^{2}]$%
, where $k_{0}$ is the strength of the inter-chain coupling. The Hamiltonian
of each chain can be written as%
\begin{equation}
H_{M}=\sum_{i=1}^{N}\left[ \frac{p_{M,i}^{2}}{2m}+\frac{k}{2}\left(
x_{M,i+1}-x_{M,i}\right) ^{2}+\frac{V}{(2\pi )^{2}}\left[ 1-\cos (2\pi
x_{M,i})\right] \right] ,
\end{equation}%
with $x_{M,i}$ and $p_{M,i}$ denote the displacement from equilibrium
position and the conjugate momentum of the $i$th particle in chain $M$,
where $M$ stands for $A$ or $B$. $N$ is the number of the particles in $A$
or $B$-chain. $m$ is the mass of the particle. The parameters $V$ and $k$
are the strength of the nonlinear external potential and the intra-chain
interaction for the FK lattice, respectively. We set the masses of all the
particles be unit and use fixed boundaries, $x_{M,N+1}=x_{M,0}=0$. The first
and the last particles of $A$-chain and $B$-chain are connected to heat
baths. The temperature of the left and right heat baths is respectively $%
T_{L}=0.25$ and $T_{R}=0.10$. The temperature used here and in the following
numerical computations is dimensionless. It is connected with the true
temperature $T_{\theta }$ of the materials through the relation \cite{Lan}: $%
T_{\theta }=(m\omega _{\theta }^{2}\phi ^{2}T)/k_{B}$, here $m$ is the mass
of the particle and $\phi $\ is the period of external potential. $\omega
_{\theta }$ is the oscillating frequency. $k_{B}$ is the Boltzmann constant.

In our simulations we use Langevin thermostat and integrate the equations of
motion by using the 4th-order Runge-Kutta algorithm \cite{Press}. We chose a
step size of the simulation $\bigtriangleup t=0.005$ and averaging over $%
2\times 10^{9}$ time steps. We have checked that our results do not depend
on the particular thermostat realization (for example, Nose-Hoover
thermostat \cite{Nose}). The local temperature is defined as $%
T_{i}=\left\langle p_{i}^{2}\right\rangle $, $\langle $ $\rangle $ means
time average. The local heat flux in double chains is defined as $%
j_{i}=k\langle p_{A,i}(x_{A,i}-x_{A,i-1})\rangle $ $+k\langle
p_{B,i}(x_{B,i}-x_{B,i-1})\rangle $ $+k_{0}\langle
p_{A,i}(x_{A,i}-x_{B,i})\rangle $ $+k_{0}\langle
p_{B,i}(x_{B,i}-x_{A,i})\rangle $, which is derived from the equations of
motion and the expression of thermal flux $\sum \left( F_{M,i}\cdot
v_{M,i}\right) $, where $M$ stands for $A$ and $B$ \cite{Lepri1}. The
simulations are performed long enough to allow the system to reach a steady
state in which the local heat flux is constant along the double-stranded
lattice. The transport coefficient is an important quantity for
characterizing the transport mode of a thermal transport process \cite{Wang}%
. The thermal conductance evaluated as $K=Nj/\bigtriangleup T$ represents an
effective transport coefficient that includes both boundary and bulk
resistances \cite{Lepri1}.

\textbf{III. RESULTS AND DISCUSSION}

The inter-chain interaction is compare with the intra-chain interaction in
the thermal conductance and the phonon spectra of the particles. Figure 2
shows the dependence of thermal conductance on the intra-chain interaction
for the system size $N=64$. For various nonlinear external potentials $V=1.0$
and $10.0$ as well as two kinds of inter-chain couplings $k_{0}=0.1$ and $%
1.0 $, the thermal conductance monotonously increases with the intra-chain
interaction. This result is good agreement with that of the single-chain FK
lattice, which is also clearly confirmed by the analytical calculations
based on the self-consistent phonon theory \cite{Shao}.
\begin{figure}[htbp]
\begin{center}\includegraphics[width=8cm,height=6cm]{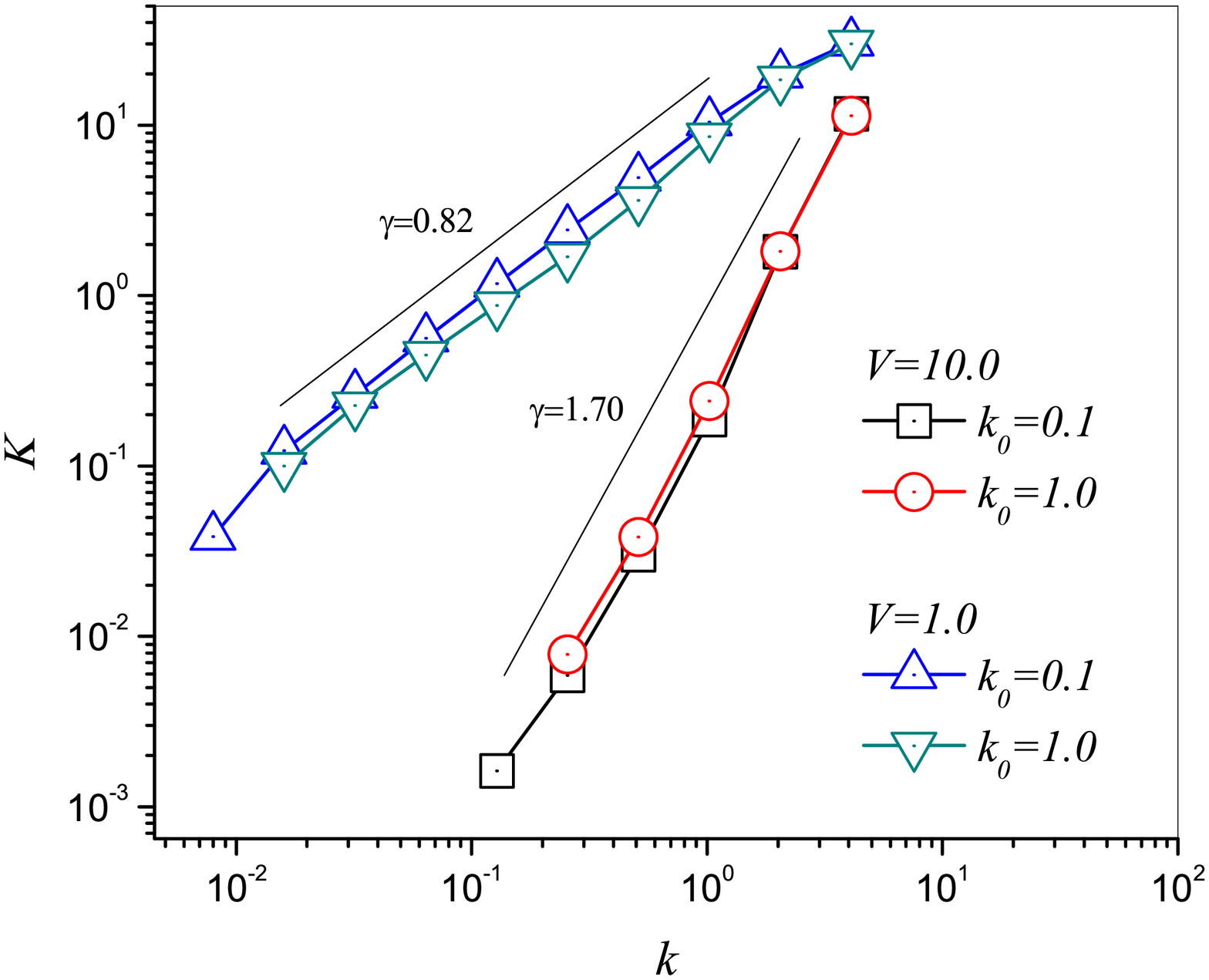}
  \end{center}
  \caption{Thermal conductance as a
function of the intra-chain interaction for various values of $V$ and $k0$
at a system size of $N=64$.}
   \label{}
\end{figure}

Thermal conductivity dependence on $k$ and $V$ can be explained by the
self-consistent phonon (SCP) theory. The SCP theory has been applied to deal
with the nonlinear Morse on-site potential for the DNA denaturation \cite%
{Dauxois}. The SCP theory can be taken into consideration the asymmetric
heat transport in the nonlinearity lattices \cite{BHu}.

Hamiltonian of FK model can be approximated an effective linear Hamiltonian
by the SCP theory as \cite{Dauxois}
\begin{equation}
H=\sum_{i=1}^{N}\left[ \frac{p_{i}^{2}}{2m}+\frac{k}{2}\left(
x_{i+1}-x_{i}\right) ^{2}+\frac{f}{2}x_{i}^{2}\right] ,
\end{equation}%
where the effective harmonic potential coefficient $f$ is obtained from the
self-consistent equation,

\begin{equation}
f=\frac{\pi ^{2}V}{a^{2}}\exp \left[ -\frac{\pi ^{2}k_{B}T}{2a^{2}\sqrt{%
f\left( 4k+f\right) }}\right] .
\end{equation}%
So, the spectrum of effective phonons is $\hat{\omega}_{k}^{2}=f+4k\sin ^{2}%
\frac{{k}_{i}}{2}.$

As reported in Ref.\cite{Shao}, using Debye formula we can get the thermal
conductivity as

\begin{equation}
K\propto \frac{ck^{3/2}}{V^{2}}\int_{0}^{2\pi }\frac{\sin ^{2}{k}_{i}}{%
(4\sin ^{2}\frac{{k}_{i}}{2}+\frac{f}{k})^{3/2}}dk_{i},
\end{equation}%
\ where $k_{i}$ is the wave vector. For the sake of simplicity, $c$ is
assumed a constant, from the equation, we can obtain the relationship of $K$
vs $V$ and $k$ as below:
\begin{equation}
K\propto \frac{k^{3/2}}{V^{2}}.
\end{equation}%
Thus it can be clearly seen that $K$ increases with $k$ as a scaling law $K%
\symbol{126}k^{\gamma }$, which is confirmed by numerical results as well as
analytical results \cite{Shao}. For single FK chain, the exponent $\gamma $
is 3/2 and is independence of the on-site potential. However, for double FK
chains, as shown in Fig.2, $\gamma $ changes with the on-site potential $V$
obviously. When $V$ increases, the exponent $\gamma $ will decreases. It can
be seen that the intra-chain interaction performs a positive influence on
thermal conductance not only for the single-chain lattice but also for the
double-chain lattice.

When we go to the double-stranded lattice and consider the inter-chain
interaction, we will find some interesting and surprising results. As shown
in Fig.3, in the case of a weak nonlinear external potential $V=1.0$ or $5.0$%
, the thermal conductance decreases with the increasing of the inter-chain
interaction. On the contrary, when the nonlinear external potential goes to
a higher value $10.0$ or $12.0$, the thermal conductance monotonously
increase with the inter-chain interaction just as with the intra-chain
interaction. Due to the universe finite size effect in low-dimensional
systems, we consider more system size $N=256$ and $512$, as shown in
Figs.3(b) and 3(c), the thermal conductance as a function of the inter-chain
interaction mentioned above is still invariant. This indicates that this
kind of anomalous heat conduction induced by the inter-chain interaction is
independent of the system size. 
\begin{figure}[htbp]
\begin{center}\includegraphics[width=8cm,height=6cm]{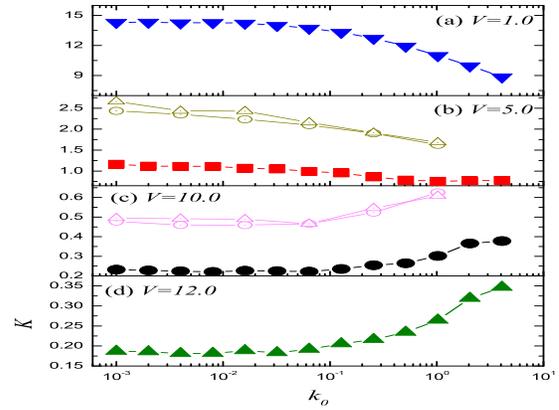}
  \end{center}
  \caption{The dependence of thermal conductance on inter-chain interaction for various
nonlinear external potential $V=1.0,5.0,10.0$ and $12.0$. The system size is
$N=64$ for full filled point ((a)blue, (b)red, (c)black and (d)green), $256$
for circle and $512$ for triangle ((b), (c)). The remaining parameter is $%
k=1.0$.}
   \label{}
\end{figure}

\begin{figure}[htbp]
\begin{center}\includegraphics[width=8cm,height=6cm]{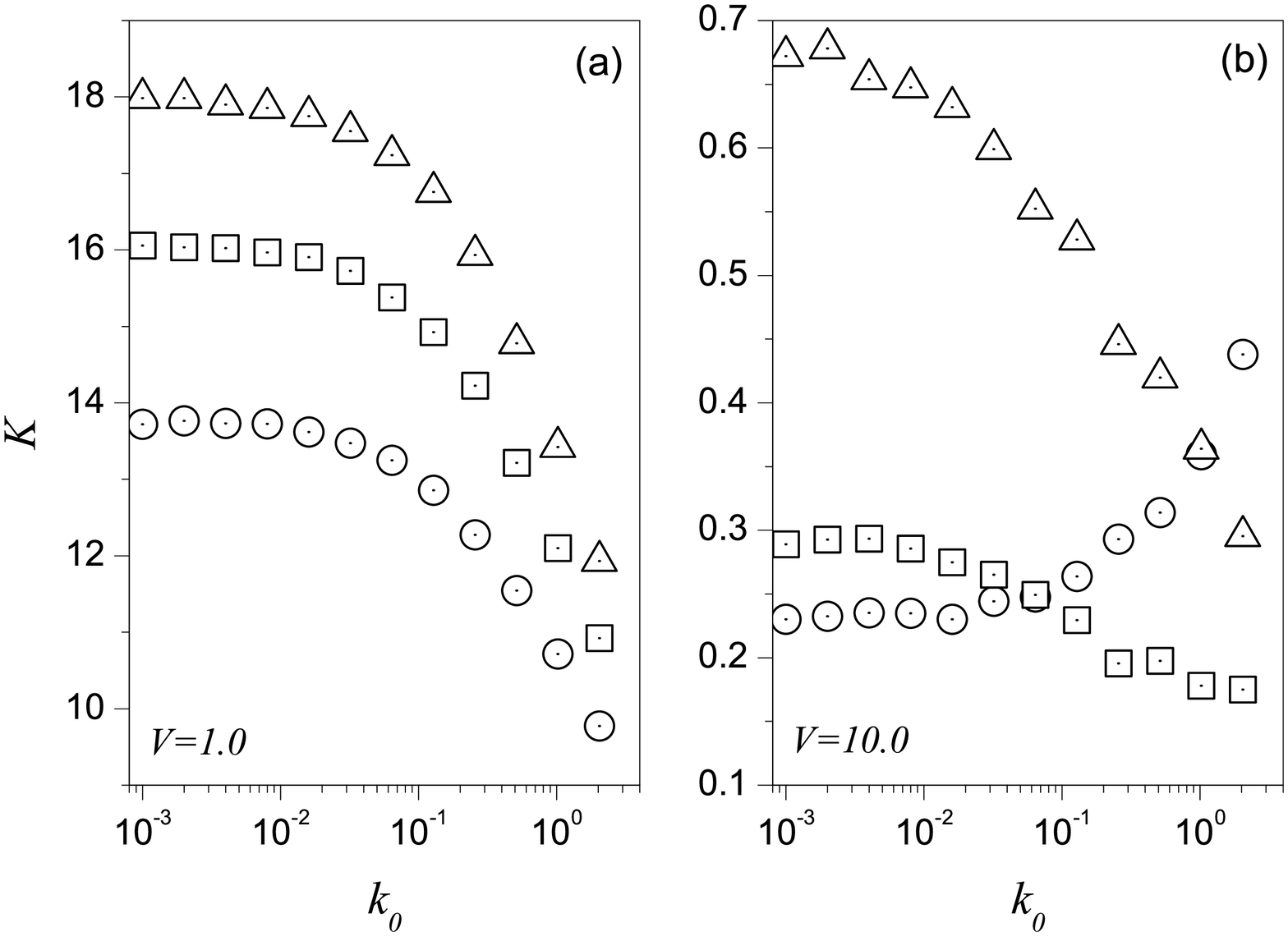}
  \end{center}
  \caption{Thermal conductance changes with the inter-chain interaction for
various temperatures $T_{0}=0.35$ (triangle), $0.25$(square), $0.15$%
(circle). }
   \label{}
\end{figure}

We also like to discuss the temperature dependence of the thermal
conductance. Here we set the temperature of the system as $T_{0}$. The heat
baths are respectively $T_{L}=T_{0}+dT$ and $T_{R}=T_{0}-dT$, where $dT$ is
0.05. In the case of weak on-site potential, as shown in Fig.4(a), the
thermal conductance decreases with the increasing of inter-chain
interaction, which is independence of the temperature. In the case of strong
on-site potential, however, the relationship between the thermal conductance
and the inter-chain interaction depends on the temperature of the system. As
shown in Fig.4(b), the thermal conductance increases with the increasing of
inter-chain coupling at low temperature ($T_{0}=0.15$); On the contrary, the
thermal conductance decreases with the increasing of inter-chain coupling at
high temperature ($T_{0}=0.25$ and $0.35$).

Firstly, we apply the transition of the thermal transport mode to interpret
the different thermal transport properties between intra and inter-chain
interaction. For the intra-chain coupling, figures 5a and 5b show the
temperature profile at different intra-chain couplings. As also shown in the
inset of Figs.5c and 5d, the thermal conductivity is finite in the diffusive
region and infinite in the ballistic region. Diffusive and ballistic regions
are respectively corresponding to large and small temperature differences.
When the intra-chain coupling increases, the temperature different in the
chain decreases, which indicate the thermal transport mode exhibits a
transition from the diffusive to the ballistic transport. Ballistic
transport means less collision of phonon and then the thermal current will
increase. Therefore, it is easily understand why the intra-chain coupling
always enhances the thermal current in the chain. For the inter-chain
coupling, as shown in Fig.5c, the inter-chain coupling cannot change the
temperature profile of the chain in the case of weak nonlinear potential.
However, Figure 5d displays the thermal transport mode exhibits a transition
from the diffusive to the ballistic transport when inter-chain coupling
increases in the case of strong nonlinear potential. The strong inter-chain
interaction refers to strong on-site potential or large effective mass of
the particle, which performs negative effects on the heat current. The
competitive effects of strong on-site potential (negative effect) and the
ballistic transport (positive effect) determine the influence of inter-chain
coupling on thermal conductance. In the case of weak nonlinear potential,
the transport mode is fixed and then the thermal conductance decreases with
the inter-chain interaction increases. In the case of strong nonlinear
potential, the on-site potential is strong enough; therefore the transition
from diffusive to ballistic transport will increase the heat current.
\begin{figure}[htbp]
\begin{center}\includegraphics[width=8cm,height=6cm]{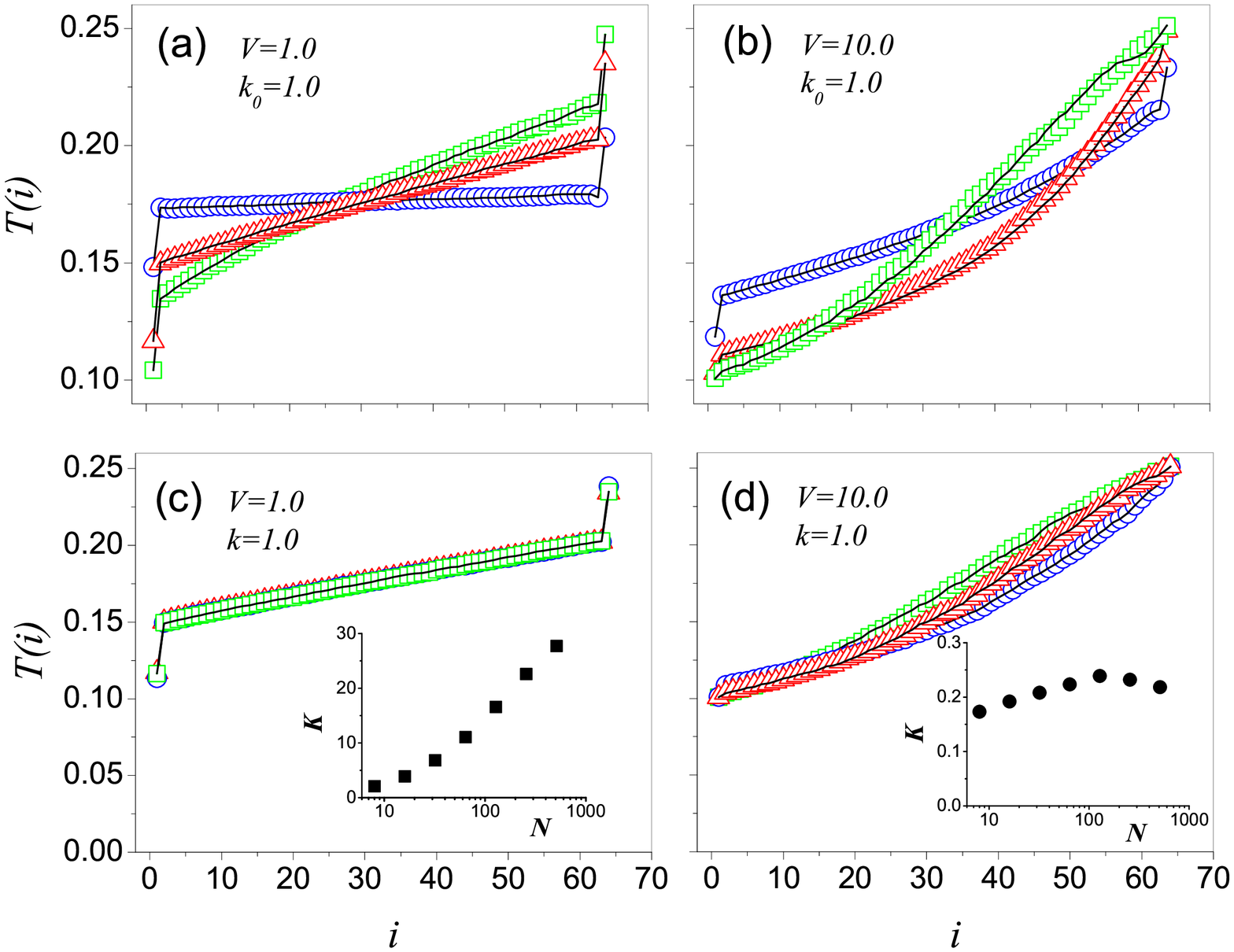}
  \end{center}
  \caption{Temperature profiles along A-chain
(point) and B-chain (line) at various intra and interchain interactions. In
(a) and (b), the parameter $k=0.1$ (Green square),$0.5$(Red triangle),$2.0$%
(Blue circle); In (c) and (d) the parameter $k_{0}=0.01$ (Green square),$0.1$%
(Red triangle),$1.0$(Blue circle). Inset: the thermal conductivity changes
with the system size for $k=1.0$ and $k_{0}=1.0.$}
   \label{}
\end{figure}

Secondly, we give another interpretation from the view of the phonon
spectra. As shown in Fig.6a, when the intra-chain interaction increases, the
band width, namely the frequency range of the mid-particle of A-chain,
expands. The phonon spectra of one particle can easily match the other,
therefore the heat current increases. We have investigated many particles
along the chain and observed similar results for other particles. However,
as illustrated in Fig.6b, when the inter-chain interaction increases, the
phonon spectra split into two branches when spreading. Obviously,
inter-chain interaction can induce another branch of oscillation mode, which
is of high oscillating frequency. The phonon is a Boson whose energy is
proportion to the frequency from equation $E=\hbar \omega $. And then the
phonon with high frequency can easily overcome the on-site potential to
transmit along the chain. On the other hand, the phonon with high frequency
has more opportunities to collide with each other, which means the thermal
conductance decreasing. The competitive effects as mentioned above are also
observed in the change of phonon spectra induced by the inter-chain
interaction.
\begin{figure}[htbp]
\begin{center}\includegraphics[width=8cm,height=6cm]{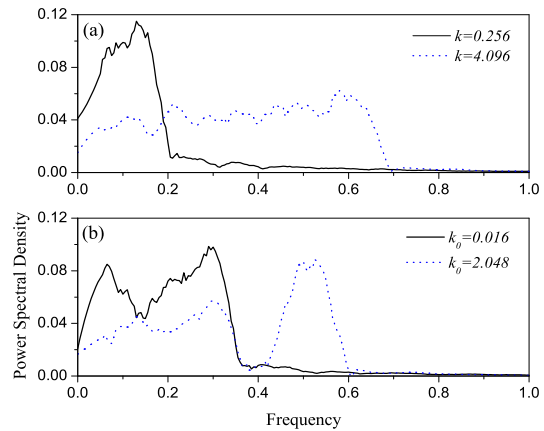}
  \end{center}
  \caption{Power spectral
density of the particle at the middle of $A-$chain for different intra and
interchain interaction with the parameter values $N=64,V=1.0$, and (a) $%
k_{0}=0.1,$(b) $k=1.0$.}
   \label{}
\end{figure}

\textbf{IV. CONCLUSIONS}

In summary, we have performed extensive numerical simulations of thermal
conduction in a double-stranded FK lattice. It is reported that the
interactions of the intra and inter-chain exhibit different thermal
transport properties. In any case, the intra-chain interaction always
upgrades the thermal conductance. The inter-chain interaction, however, has
both positive and negative effects on the thermal conductance: the positive
effect at the strong nonlinear potential and the negative effect at the weak
nonlinear potential. Moreover, the changes of the phonon spectra of the
particles, which are induced by intra-chain interaction, are quite different
from those induced by inter-chain interaction. It is suggested that a simple
coupling between the particles, which is a common point for both intra-chain
and inter-chain interaction, can develop various thermal transport phenomena
under different situations. Although our model based on one dimensional
lattice is insufficient to give a realistic description of real DNA and
polymer systems, our result will provide a new physical view of DNA,
polymer, nanomaterials and others structure, which are similar to
double-stranded lattice.

\textbf{ACKNOWLEDGEMENTS}

I really appreciate Prof. Bambi Hu for giving me helpful hand and for the
usage of their computing cluster while I was in an extremely difficult
situation. I would also like to thank members of the Centre for Nonlinear
Studies in Hong Kong Baptist University for valuable discussions. This work
was supported in part by grants from the Jinan University Faculty Research
Grant FRG (Grant No.50624006), the Fundamental Research Funds for the
Central Universities, JNU (Grant No.21609305) and the National Natural
Science Foundation of China (Grant No.10875178).

\end{document}